\documentclass[pra,reprint,doublecolumn,superscriptaddress, nobalancelastpage,dvipsnames]{revtex4-2}

\usepackage[utf8]{inputenc}
\usepackage[T1]{fontenc}
\usepackage{bm}

\usepackage[dvipsnames]{xcolor}

\usepackage{tikz}
\usepackage[sort&compress]{natbib}
\usepackage{enumitem,kantlipsum}
\usepackage{graphicx}
\usepackage{verbatim}
\usepackage{float}
\usepackage{sidecap}
\sidecaptionvpos{figure}{c}
\usepackage{lipsum}
\usepackage{capt-of}
\usepackage{setspace}
\usepackage{dcolumn}
\usepackage{bm}
\usepackage{amssymb}
\usepackage{amsmath}
\usepackage{mathtools}
\usepackage{physics}
\usepackage{multirow}
\usepackage[a4paper]{geometry}
\newgeometry{top=1.25cm,right=1cm,left=1.5cm,bottom=1.25cm}
\usepackage[colorlinks=true, urlcolor=blue, pdfborder={0 0 0}]{hyperref}
\hypersetup{%
linkcolor=blue,
citecolor=BrickRed
}

\draft

\begin{document}

\title{\Large{\textbf{Towards quantum enhanced adversarial robustness in machine learning}}}

\author{Maxwell T. West} \email{westm2@student.unimelb.edu.au}  \affiliation{School of Physics, The University of Melbourne, Parkville, 3010, VIC, Australia}

\author{Shu-Lok Tsang}
\affiliation{School of Computing and Information Systems, The University of Melbourne, Parkville, 3010, VIC, Australia}

\author{Jia S. Low}
\affiliation{School of Computing and Information Systems, The University of Melbourne, Parkville, 3010, VIC, Australia}

\author{Charles D. Hill}
\affiliation{School of Physics, The University of Melbourne, Parkville, 3010, VIC, Australia}
\affiliation{School of Mathematics and Statistics, The University of Melbourne, Parkville, 3010, VIC, Australia}

\author{Christopher Leckie}
\affiliation{School of Computing and Information Systems, The University of Melbourne, Parkville, 3010, VIC, Australia}

\author{Lloyd C.L. Hollenberg} \affiliation{School of Physics, The University of Melbourne, Parkville, 3010, VIC, Australia}
\affiliation{Center for Quantum Computation and Communication Technologies, The University of Melbourne, Parkville, 3010, VIC, Australia}

\author{Sarah M. Erfani}
\affiliation{School of Computing and Information Systems, The University of Melbourne, Parkville, 3010, VIC, Australia}

\author{Muhammad Usman} \email{musman@unimelb.edu.au}  \affiliation{School of Physics, The University of Melbourne, Parkville, 3010, VIC, Australia}
\affiliation{Data61, CSIRO, Clayton, 3168, VIC, Australia}

\maketitle

\onecolumngrid

\noindent
\textcolor{black}{
  \normalsize{\textbf{Machine learning algorithms are powerful tools for data driven tasks such as image classification and feature detection, however their vulnerability to adversarial examples - input samples manipulated to fool the algorithm - remains a serious challenge. The integration of machine learning with quantum computing has the potential to yield tools offering not only better accuracy and computational efficiency, but also superior robustness against adversarial attacks. Indeed, recent work has employed quantum mechanical phenomena to defend against adversarial attacks, spurring the rapid development of the field of quantum adversarial machine learning (QAML) and potentially yielding a new source of quantum advantage. Despite promising early results, there remain challenges towards building robust real-world QAML tools. In this review we discuss recent progress in QAML and identify key challenges. We also suggest future research directions which could determine the route to practicality for QAML approaches as quantum computing hardware scales up and noise levels are reduced.}}} \\ 

\twocolumngrid

\begin{figure*}[ht]
 \begin{center}
 \includegraphics[width=19 cm]{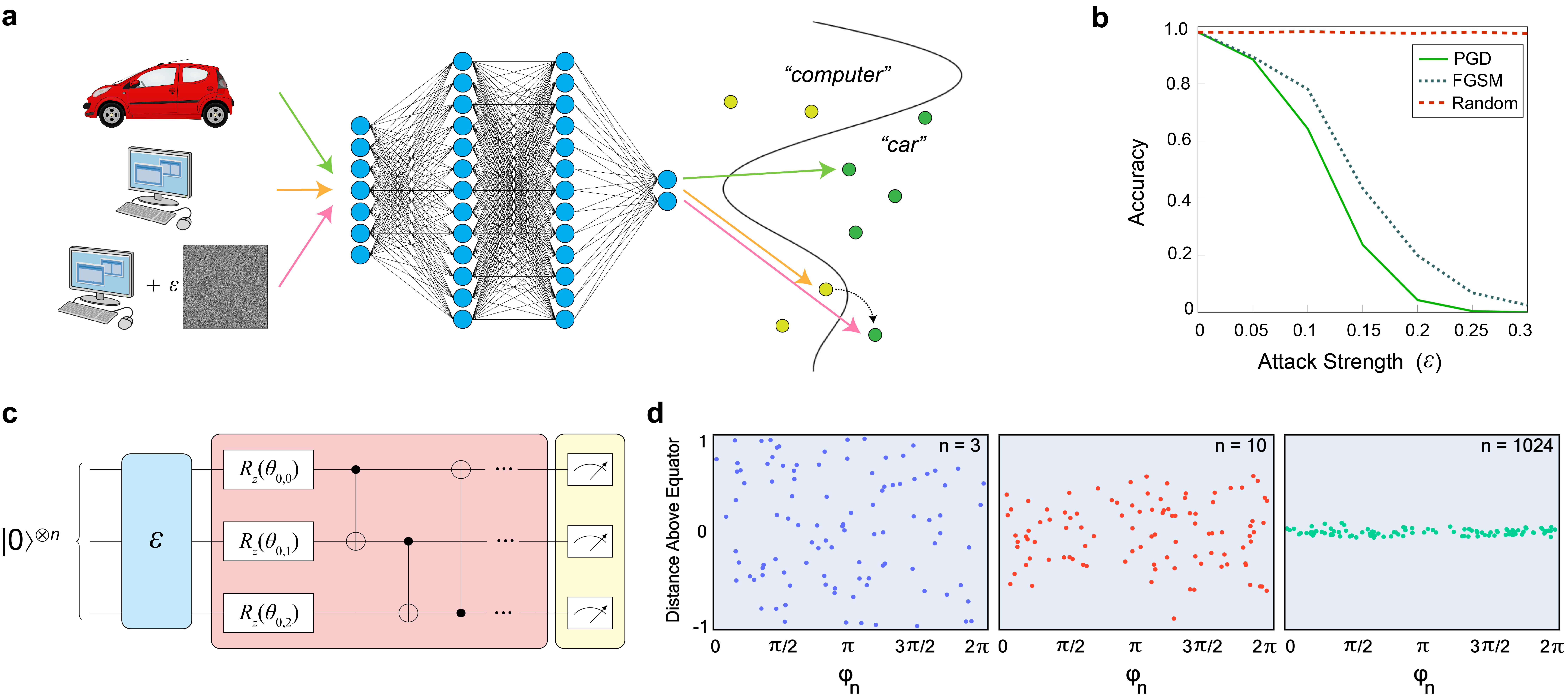}
   \caption{\label{fig:1}\textbf{Adversarial machine learning.} \textbf{a.} Common machine learning methods can be thought of as 
   performing an embedding from an input space of possible examples to an internal space used for classification. A classifier 
   will construct a \textit{decision boundary} in the classification space, with inputs classified according to  on which side 
   of the boundary they fall. As depicted schematically in the figure, inputs which are mapped close to the decision boundary 
   are subject to the possibility of being misclassified under a small adversarial perturbation, despite potentially being all 
   but indistinguishable from the original clean data sample. 
   \textbf{b.} The accuracy of a convolutional 
   neural network on the MNIST \cite{726791} dataset of images of handwritten digits
   in the presence of perturbations generated by the PGD attack \cite{madry2017towards}, the FGSM attack \cite{goodfellow2014explaining}, and randomly. The strength of the 
   attack is the maximum amount by which an individual pixel may be changed (the pixels take values in $[0,1]$). The mere magnitude 
   of the changes is insufficient to cause misclassification, as can be seen by the robustness to randomly generated perturbations, 
   but the targeted attacks are capable of conclusively deceiving the network. 
   \textbf{c.} A typical quantum classifier consists of three main components: first the data is encoded into a quantum state (blue), then 
   fed through multiple layers of parameterised rotations and entangling gates (red), and finally a measurement is performed to 
   determine the output of the model (yellow). 
   \textbf{d.} The height above the equator and the azimuthal angle $\varphi_n$ of 100 points  sampled uniformly randomly from unit 
   spheres  $ \mathbb{S} ^{  n - 1} \subseteq   \mathbb{ R} ^{  n}  $ for  $  n \in \{3,\ 10,\ 1024 \}$. While $\varphi_n$ 
   is distributed uniformly on the interval $[0,2\pi)$ independently of $n$,  randomly selected points are found exponentially 
   (in the dimension $n-1$ of the sphere) close to the equator. By symmetry there is nothing special about the chosen equator;  
   counter-intuitively, points on a high dimension or sphere concentrate about \textit{any} equator. Moreover they will 
   cluster around the boundary of any set of measure at least 1/2, which necessarily includes the decision boundary of any 
   binary classifier on $ \mathbb{S} ^{  n - 1}$.   
   Such a classifier would then have significant adversarial vulnerability; almost all of the points in $ \mathbb{S} ^{  n - 1}$ 
   lie close to the decision boundry, and are therefore vulnerable to being adversarially perturbed over it.
   }
 \end{center}
 \end{figure*}

\noindent
Machine learning (ML) algorithms are ubiquitous nowadays and underpin the vast majority of autonomous and robotic systems, including 
those deployed in security applications such as facial recognition, data classification, surveillance, and security systems 
for military applications~\cite{lecun2015deep}. In these settings, the robustness of ML algorithms is of critical importance, with any vulnerability to data manipulation potentially posing a serious security threat. This has instigated a major new subfield within machine learning, namely adversarial machine learning. Adversarial ML is concerned with the process of generating inputs which will be misclassified by a targeted ML system 
(typically a neural network), despite being only perturbed by a small amount from an initial, correctly classified input 
~\cite{biggio2013evasion, szegedy2013intriguing,10.1145/2046684.2046692, kurakin2016adversarial} (see Figure~\ref{fig:1}).  
While modern neural networks are generally resilient to minor random perturbations of their inputs,  they can be extremely susceptible to nonrandom,  carefully crafted ones as shown in Figure~\ref{fig:1}(b). In the case of high resolution image  classification,  even state-of-the-art 
convolutional neural networks can be fooled by adding  to a clean image perturbations which are so small they are completely 
imperceptible to human eyes~\cite{szegedy2013intriguing}, or possibly consisting of a change to only a single pixel~\cite{su2019one}.   
The surprising brittleness of such powerful classifiers  has been  intensively studied in recent years,  with increasingly 
sophisticated methods of attacking~\cite{athalye2018obfuscated, goodfellow2014explaining, kurakin2018adversarial, eykholt2018robust, carlini2017adversarial} (\textit{i.e.}  generating adversarial examples) and defending~\cite{DBLP:journals/corr/abs-2001-03994, madry2017towards, 10.1145/3134599, miller2020adversarial} neural networks developed.  
\\ \\
\noindent
In a world where security sensitive tasks are beginning to be 
outsourced to ML frameworks, it is  imperative to fully understand the nature of the mechanism by which neural 
networks may be tricked by seemingly innocuous examples which are all but indistinguishable from genuine data \cite{ilyas2019adversarial}. This need is 
heightened by the recent discoveries of adversarial attacks originating not from applying such perturbations digitally, but 
rather carrying out attacks in the physical world~\cite{eykholt2018robust, 10.1145/2976749.2978392, eykholt2018robust}, for example by applying stickers 
in an adversarial fashion to road signs~\cite{eykholt2018robust}, with potentially serious implications for the reliability of self 
driving cars. While for now the long-term resolution of the problems of adversarial examples remains unclear, recently attention 
has also been turning to how quantum machine learning (QML) ~\cite{qml} will fare against adversarial 
attacks~\cite{lu2020quantum, PhysRevA.101.062331, du2021quantum, guan2021robustness, weber2021optimal, ren2022experimental, liao2021robust, kehoe2021defence, west2022benchmarking}.

\begin{figure*}
\begin{center}
\includegraphics[width=16 cm]{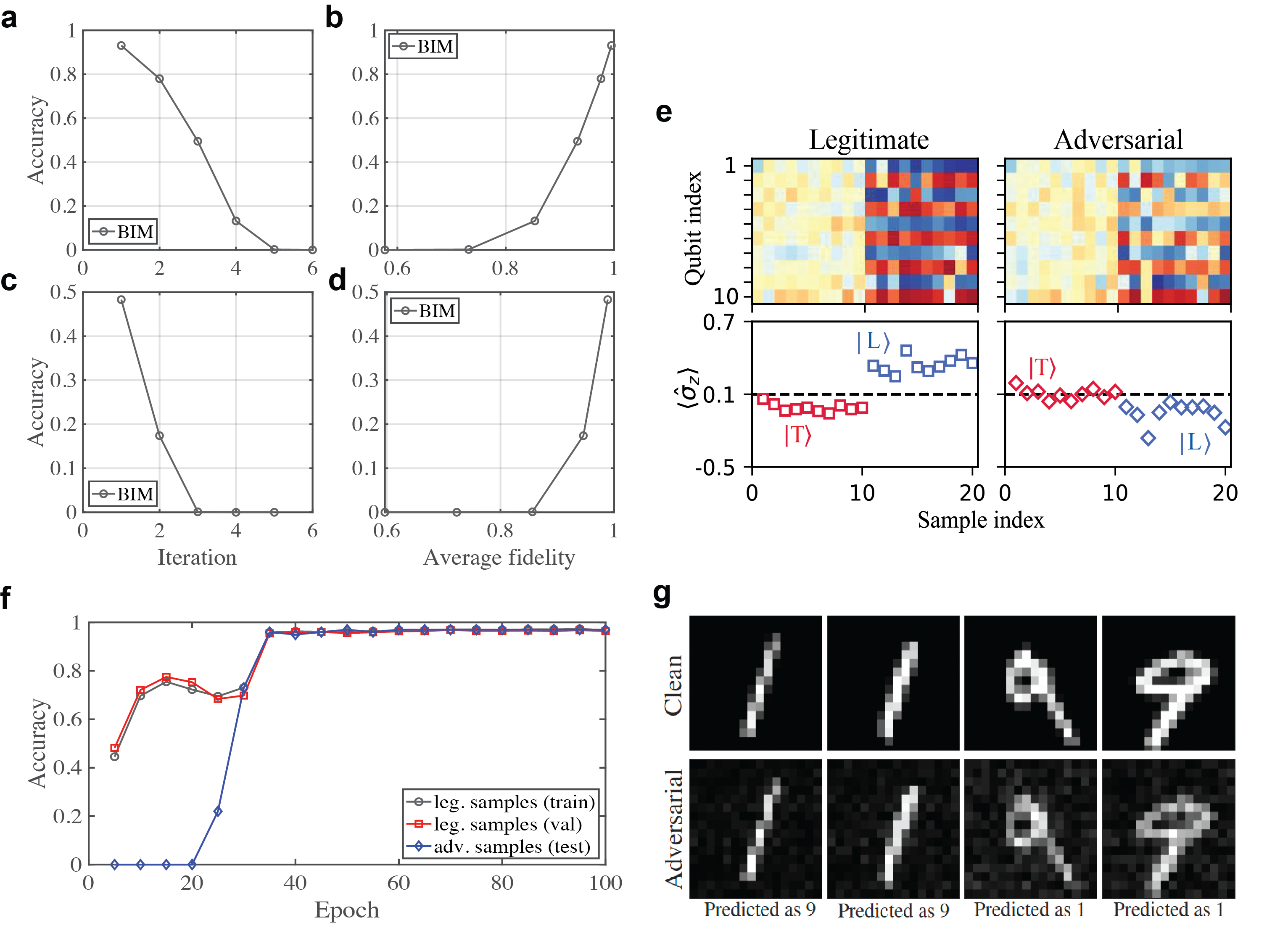}
\captionof{figure}{\textbf{Attacking and Defending Quantum Classifiers} 
  \textbf{a.} The accuracy of a binary QVC trained to distinguish between the digits 1 and 9 on adversarial examples as a function 
  of the number of image-modifying attack iterations (of step size 0.1) as obtained in Ref \cite{lu2020quantum}. \textbf{b.} 
  The average fidelity between the adversarial examples and their corresponding clean images, which remains 
  high even as the accuracy drops to zero. \textbf{c,d.} Similar to \textbf{a,b} except for a classifier trained to distinguish 
  between 1, 3, 7 and 9. 
  \textbf{e.} The system described by this Hamiltonian undergoes a phase transition from 
  thermal $(\ket{T})$ to localised $(\ket{L})$ phases at a certain critical point; determining the phase of a given state 
  is the classification problem considered by Ren \textit{et al}~\cite{ren2022experimental}.
  The  predictions 
  of the classifier on legitimate and adversarial data, as determined by the sign of $\expval{\sigma_{z}}$. 100\% 
  accuracy is obtained on the clean data, but only 25\% on the adversarial data, despite their clear visual 
  similarities. 
  \textbf{f.} The accuracy obtained on legitimate and adversarial samples by Lu \textit{et al} \cite{lu2020quantum} during adversarial training. 
  The initial oscillations observed are a result of the competition between the minimisation and maximisation loops of adversarial training.
  After adversarial training near perfect accuracy is obtained on the adversarial samples (\textit{cf} parts \textbf{a-d}).
  \textbf{g.} Clean and corresponding adversarial examples generated by attacking the QVC of Lu \textit{et al}. 
  In all cases the clean image is correctly classified and the adversarial image incorrectly classified, despite their clear similarities.
  Panels \textbf{a-d} are reprinted from ref. \cite{lu2020quantum}, American Physical Society. 
  Panel \textbf{e} is reprinted from ref. \cite{ren2022experimental}, Nature publishing group. 
  Panels \textbf{f-g} are reprinted from ref. \cite{lu2020quantum}, American Physical Society.}
\label{fig:attack}
\end{center}
\end{figure*}

\noindent
The incredible advancements in quantum computing over the last few years, both on the hardware and software fronts, have led to quantum versions of most common ML algorithms~\cite{beer2020training,havlivcek2019supervised,dallaire2018quantum,lu2014quantum,romero2017quantum,kehoe2021defence}, giving birth to a new field of QML~\cite{qml}, with the anticipation that QML tasks may be amongst the first to demonstrate quantum advantage on near-term quantum computers \cite{riste2017demonstration}. In addition to offering new ways of studying classical data, quantum methods may also be naturally used to study quantum data, with advantages for doing 
so beginning to emerge even on the noisy hardware available today \cite{huang2022quantum}. Although the majority of the focus in searching for quantum advantage through QML has been on the question of algorithmic speed-ups, the existence of adversarial examples provides another possible route to advantage -- enhanced adversarial robustness. Indeed, the field of quantum adversarial machine learning (QAML) has recently gained much attention~\cite{lu2020quantum, PhysRevA.101.062331, du2021quantum, guan2021robustness, weber2021optimal, ren2022experimental, liao2021robust, kehoe2021defence,west2022benchmarking}. A typical QML framework based on a variational circuit approach is shown in Figure~\ref{fig:1} (c), which has been the subject of recent studies to judge quantum-enabled robustness of ML tools. Early work~\cite{PhysRevA.101.062331} has suggested that QML methods may be highly vulnerable to adversarial attacks as a result of a counter-intuitive geometric property (the concentration of measure phenomenon (COMP) \cite{Ledoux2001TheCO}) of the Hilbert spaces they employ for classification, independent of the precise details of the classifier. Moreover, as adversarial examples are not sampled randomly from the input data distribution, but rather carefully constructed, the promising recent work on generalisation bounds for QML~\cite{caro2021encoding, carogeneralization2022, banchigeneralization2021} cannot provide guarantees of adversarial robustness. While the COMP suggests that undefended quantum classifiers may be highly vulnerable to attack, recent work has shown that the unique properties of quantum information processing may, on the other hand, offer an opportunity to design new methods of defending against adversarial attacks~\cite{du2021quantum,weber2021optimal}, setting up a fascinating battle between attack and defence in the field of QAML.
\\ \\
\noindent
In this paper, we have provided a succinct review of the recent literature on QAML, examining the progress in both offensive and defensive measures. We have identified some of the key challenges and potential future research directions in each of the key components of the Q(A)ML pipeline: attacks, data encoding, network architecture, and noise mitigation.
\\ \\
\Large{\textbf{State of the Art}}
\normalsize
\\ \\
\noindent
\textbf{Attacking Quantum Classifiers.} 
The adversarial vulnerability of quantum classifiers stems from a counter-intuitive property of their underlying Hilbert spaces~\cite{PhysRevA.101.062331}. Specifically, 
these Hilbert spaces, into which the inputs are mapped, have the property that their points strongly cluster around the decision boundaries of the classifiers (see Figure~\ref{fig:1}(d)). This has significant implications for adversarial robustness of the classifiers, as points which are close to a decision boundary are precisely the points which are 
susceptible to adversarial perturbation (see Figure \ref{fig:1}(a)). Specifically,
the COMP implies that the distance from a (Haar) randomly sampled point in the Hilbert space to the nearest adversarial example vanishes as $\mathcal{O}\left(2^{-n_\mathrm{qubits}}\right)$ \cite{PhysRevA.101.062331}, implying a drastic vulnerability even for relatively modest numbers of qubits. Moreover, given $k$ independent quantum classifiers, one can find universal adversarial examples which deceive all classifiers with only a perturbation of strength $\mathcal{O}\left(\log (k)2^{-n_\mathrm{qubits}}\right)$ \cite{gong2022universal}. 
In practical ML applications, however, one 
will not be sampling randomly from the entire space, but rather focusing on some small subset consisting of the 
encoded states of meaningful examples \cite{PhysRevA.101.062331, liao2021robust, larose2020robust, dallaire2018quantum, creswell2018generative},
which need not necessarily exhibit concentration of measure. In the commonly used phase encoding scheme $\boldsymbol{x}\mapsto\otimes_{i=1}^n (\cos(x_i)\ket{0}+\sin(x_i)\ket{1})$, for example, the necessary perturbation strength scales only as $\mathcal{O}\left(1/\sqrt{n_\mathrm{qubits}}\right)$~\cite{liao2021robust}. The therefore considerable impact of the type of states being classified on adversarial robustness implies both that the impact on robustness is another factor to consider in the choice of a data encoding scheme in QML, and that thorough empirical studies on common datasets are needed to investigate 
the performance of adversarial attacks on quantum classifiers in practice.
\\ \\
\noindent
Such an empirical study has recently been conducted by Lu \textit{et al} \cite{lu2020quantum}, in which
attacks were carried out on a standard quantum variational classifier (QVC) (see Figure~\ref{fig:1}(c)) using typical attacks \cite{goodfellow2014explaining,kurakin2018adversarial,madry2017towards,chen2017zoo} 
developed in the classical setting, and both classical (MNIST)~\cite{726791} and quantum datasets~\cite{jiang2019adversarial}. 
Some typical adversarial examples generated by attacking a QVC trained to classify MNIST data
are shown in Figure \ref{fig:attack}(g).  In each case the QVC can be tricked into misclassifying an image that is very similar to a 
corresponding, correctly classified clean image. The average accuracy achieved on their calculated adversarial examples
are shown in Figure \ref{fig:attack}(a-d), providing clear empirical evidence of the 
susceptibility of quantum classifiers to adversarial perturbations in practice - even in the relatively simple case of binary classification on MNIST data. For more complicated data, where the features which need to be identified in order to facilitate accurate classification are more subtle, this susceptibility would only be expected to increase. 

\noindent
Although the simulations of Ref. \cite{lu2020quantum} have predicted a severe vulnerability of quantum classifiers 
to adversarial attacks, the true test of their vulnerability must be carried out on real quantum hardware, in the accompanying presence of 
quantum noise and hardware constraints
(the potential benefits of quantum noise in reducing adversarial vulnerability are discussed in the section
``Adversarial Robustness Through Quantum Noise'').
While the limitations imposed by the current NISQ era of 
quantum hardware make an experimental implementation of QAML a challenging task, 
nevertheless, and in an important study for the field, such an experiment has been reported very recently by Ren \textit{et al} 
\cite{ren2022experimental}. Considering both classical image data and quantum data, they succeeded in both training high fidelity QVCs 
and subsequently attacking them, producing adversarial examples which fool their classifiers despite 
retaining a high degree of visual similarity to their corresponding legitimate samples (see Figure \ref{fig:attack}(e)). 
This work marks an end-to-end experimental implementation of QAML, demonstrating that the field is concerned with
real effects which can be seen already on current noisy hardware. 
\noindent
\\ \\
\noindent
\textbf{Defending Quantum Classifiers.}
\normalsize
\noindent
The classical ML community has in recent years seen a flurry of discoveries involving an 
increasingly sophisticated series of adversarial attacks, defences, and counter-attacks 
\cite{goodfellow2014explaining, madry2017towards, athalye2018obfuscated, carlini2017adversarial, miller2020adversarial, guo2017countering, 46641, feinman2017detecting, salman2019provably,zhang2019theoretically}. While many of the classical defence strategies can be expected to automatically carry over to the quantum setting given their architecture agnostic form, recent work \cite{du2021quantum, guan2021robustness, weber2021optimal} has also begun to explore the 
possibility of defence mechanisms which rely purely on quantum phenomena. Due to the observed trend in the classical setting of seemingly strong defensive strategies 
subsequently failing against improved methods of attack, another important area of study is that of \textit{certifiable robustness}, wherein guarantees of the non-existence 
of adversarial examples within a certain distance of clean samples are derived 
\cite{lecuyer2019certified, cohen2019certified,wong2018scaling,raghunathan2018certified,10.1007/978-3-030-53288-8_2, elboher2020abstraction, DBLP:journals/corr/abs-2005-07173, kwiatkowska2019safety, du2021quantum, weber2021optimal}. We now briefly discuss three of the key defensive techniques and guarantees of adversarial robustness which have been introduced in the QAML literature, covering both strategies which are direct applications of classical techniques, and strategies which harness non-classical aspects of quantum computing.

\begin{enumerate} [leftmargin=*]

\item \textbf{Adversarial Robustness through Quantum Noise.}
\normalsize
\noindent
Quantum noise is generally regarded as a liability which poses major challenges for current quantum devices to tackle real-world applications. However, 
random noise plays a natural role in adversarial attack mitigation: the intentional injection of a small amount of randomness 
into the classification process can help to thwart adversarial attacks by ``drowning out'' any potential adversarial perturbation.
This idea has been successfully implemented in the classical setting \cite{cohen2019certified}, including a guarantee of certifiable robustness.
The pursuit of similar results in the quantum case has already begun, with promising results reported~\cite{du2021quantum, weber2021optimal}. Moreover, 
the inherently random nature of the output of a quantum circuit presents an opportunity for seeking robustness bounds even in the 
noiseless case, unlike classically where noise must be explicitly introduced.

\begin{figure*}
\begin{center}
\includegraphics[width=15 cm]{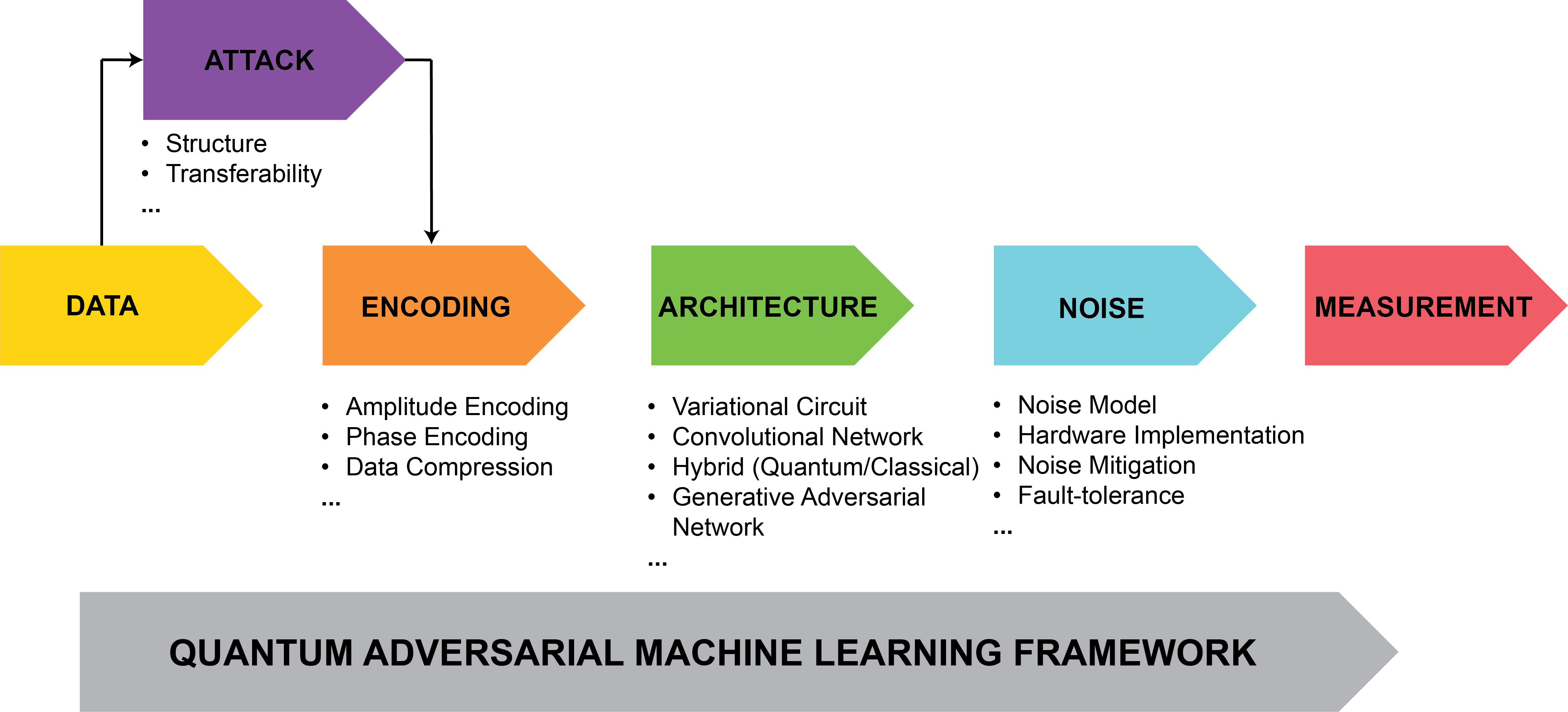}
\captionof{figure}{\textbf{Quantum Adversarial Machine Learning Framework.} A generic framework for the implementation of quantum adversarial machine learning. For each building block in the framework, we identify key areas of challenges and opportunities.}
\label{fig:qaml}
\end{center}
\end{figure*}

\noindent
On the quantum front, Ref. \cite{du2021quantum} has demonstrated that depolarisation noise in a quantum circuit  can
induce \textit{quantum differential privacy} \cite{zhou2017differential}, a natural generalisation of the 
concept of differential privacy from classical computer science \cite{dwork2008differential, lecuyer2019certified}. 
Differential privacy is a property of randomised algorithms which are ``not too sensitive'' to small changes in 
their inputs, which is desirable behaviour in the pursuit of adversarial robustness. 
In the case of Ref. \cite{du2021quantum}, the presence of the noise 
quantifiably limits the possible influence of adversarial perturbations, but the usefulness of the classifier is preserved due to its 
inherent robustness to depolarisation noise~\cite{du2021quantum}. From the insensitivity of algorithms with differential privacy to small changes in their inputs, they then 
show \cite{du2021quantum} that the quantum classifier possesses a guarantee of adversarial robustness.

\item \textbf{Certifiable Robustness of Quantum Classifiers.}
\normalsize
\noindent
A major step towards the understanding of the certifiable robustness of quantum classifiers in the face of adversarial perturbations 
has recently been made by Weber \textit{et al} \cite{weber2021optimal}, who have drawn a connection between provable robustness and 
\textit{quantum hypothesis testing} (QHT) \cite{helstrom1967detection, holevo1973statistical}. QHT is concerned with the problem of distinguishing between two quantum states as reliably as possible, an
important task  in quantum information theory. Intuitively, the result is based on the idea that if two states cannot be reliably 
distinguished even by an optimal measurement, then neither can they be distinguished by a given quantum classifier. If one of 
such a pair of states is an adversarially perturbed version of the other, then, the classifier will nonetheless assign them the 
same label, limited as it is in its ability to even tell them apart, let alone classify them differently. 
This is exactly the behaviour we seek when attempting to obtain adversarially robust classifiers. 
Relying as it does on the fundamental inability of any protocol to flawlessly distinguish between two non-orthogonal quantum states, a property of 
quantum mechanics which is absent classically, this approach constitutes another example of dealing with adversarial perturbations 
in a way which is unique to quantum machine learning. 
Moreover, the bound that they derive is provably optimal in the important case of binary classification, 
and whenever the classifier's top choice is selected with probability greater than one half \cite{weber2021optimal}.

\item \textbf{Adversarial Training.}
\normalsize
\noindent
One of the central defensive techniques in classical ML is that of \textit{adversarial training}, wherein adversarial examples are generated at training time, and included in the training set \cite{goodfellow2014explaining, DBLP:journals/corr/abs-2001-03994, madry2017towards, bai2021recent}. While such a procedure cannot yield rigorous robustness 
guarantees, adversarial training is nonetheless considered a standard benchmark technique by the classical ML community due 
to its strong results in practice \cite{DBLP:journals/corr/abs-2001-03994, bai2021recent}. Early results \cite{lu2020quantum, ren2022experimental} have also shown promising results for the use 
of adversarial training as a method for generating robustness for quantum classifiers. In Ref \cite{lu2020quantum}, for example, Lu \textit{et al} undertake 
adversarial training of their MNIST quantum classifier (see Figure \ref{fig:attack}(f)), achieving high accuracy on both the legitimate and adversarial test samples. This is in sharp 
contrast to the situation of vanilla training, where adversarial attacks had a devastating effect on the accuracy of their classifier (see Figure \ref{fig:attack}(a-d)). Despite these encouragingly successful early applications of adversarial training in the quantum setting, an important caveat is that the attacks trialled at test time were generated using the same method as the attacks on the training data; comparably high performance when the attacks at test time are of a different nature is not guaranteed \cite{bai2021recent, kang2019transfer}. Nonetheless, the further investigation of adversarial training constitutes an important and interesting future research direction within QAML. 
\end{enumerate}

\noindent
\\ \\ \\
\Large{\textbf{Challenges and Opportunities}}
\normalsize
\\ \\
\noindent  
Figure~\ref{fig:qaml} shows a generic framework for the implementation of QAML, which consists of six building blocks. Apart from the Data and Measurement units, the other four building blocks (Attack, Encoding, Architecture, Noise) of the QAML framework require substantial future research and development en-route to practicality of QAML. In this section, we discuss critical areas of development in each block, highlight some of the major open questions and propose future research directions. 
\\ \\
\noindent
Note that the development in the areas of QAML and QML is closely related, therefore some of the challenges and opportunities are shared among the two fields -- for example, efficient data encoding and overcoming hardware noise is crucial for both. Conversely, QAML and QML target distinct goals -- QML requires high learning accuracies, whereas QAML primarily focuses on robustness against attacks -- and therefore there are many features which are unique for QAML. For example, generation and evaluation of attacks is only relevant for QAML. Likewise, hardware noise may help in certain circumstances for QAML, but it is typical detrimental for QML applications. Finally, an architecture which may offer optimal robustness against adversarial attacks may not offer optimal performance for QML tasks.
\\ \\
\noindent
\textbf{ATTACK.}
\normalsize
\noindent
A surprising feature of adversarial examples in the classical setting is that of 
\textit{transferability} \cite{szegedy2013intriguing, goodfellow2014explaining}, \textit{i.e.}, adversarial examples constructed by attacking a specific 
network (a \textit{white-box} attack) tend to act as adversarial examples for other, completely independent networks. Transferability underlies the vulnerability of ML classifiers in practice; even unfamiliar networks to which an adversary does not have direct access are susceptible as the adversary may create their own ML network, attack it, and then transfer the generated examples to the target.
\\ \\
\noindent
It has been conjectured \cite{ilyas2019adversarial} that this behaviour is a result of 
different networks independently discovering the same set of highly informative, but non-robust, features in the dataset; by targeting these features an adversarial example may then fool multiple classifiers. The complicated nature of these non-robust features leads to
adversarial perturbations which are themselves highly complex,
often appearing to contain no meaningful semantic information.
Alternately, more meaningful perturbations may be obtained
by attacking networks which have been 
forced to identify more robust
features of the data, for example by undergoing adversarial training~\cite{tsipras2018robustness}. The structure of the adversarial perturbations generated by quantum attacks, on the other hand, is not yet well-understood and it is an interesting open question if they will naturally resemble those generated by attacking robust or non-robust classical networks. A recent theoretcial study, for the first time, made a surprising revelation that the adversarial perturbations generated by quantum attacks exhibit notable structure, targetting robust features \cite{west2022benchmarking}, akin to what would be in the case of explicit adversarial training in a classical setting. This allowed QAML networks to generate very strong attacks which were able to fool powerful classical networks, demonstrating a high level of quantum to classical transferability. Further work is needed to fully understand the origin of such behavior in QAML architectures. 

\begin{figure*}
\begin{center}
\includegraphics[width=19 cm]{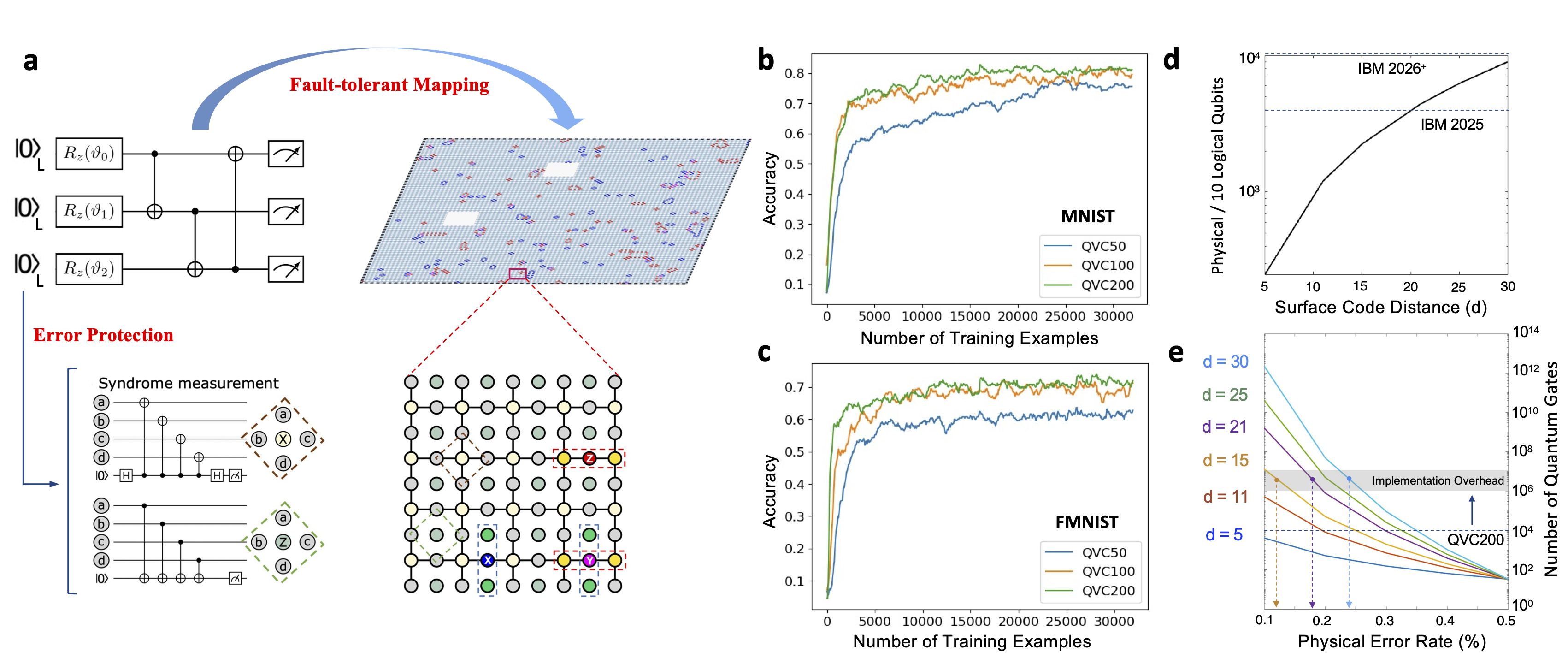}
  \captionof{figure}{\textbf{Fault-tolerant Implementation.} \textbf{a.} A conceptual illustration of the implementation of quantum machine learning networks on error-protected logical qubits. The surface code illustration is adapted from Reference \cite{gicev2021scalable}. \textbf{b,c.} Learning accuracies of quantum variational machine learning networks for two datasets: MNIST and FMNIST. The quantum networks consists of 50, 100, and 200 layers (labelled as QVC50, QVC100, and QVC200), with each layer made up of single qubit rotation gates (trainable) and controlled-Z gates. \textbf{d.} A plot of the number of physical qubits required to implement ten error protected logical qubits as a function of surface code distances (d). The horizontal dashed lines indicate number of physical qubits in IBM Quantum devices anticipated in 2025 and beyond 2026 based on the IBM Quantum roadmap. \textbf{e.} A plot of number of fault-tolerant quantum gates (circuit depth) as a function of the qubit physical error rates for various surface code distances. The shaded region indicates the number of quantum gates required for QVC200 implementation including surface code overheads (magic state distillation and synthesis). The dashed vertical lines indicate the approximate physical error rates for the fault-tolerant QVC200 implementation.}
\label{fig:FTI}
\end{center}
\end{figure*}

\noindent
The targeting of specific features of the data by adversarial attacks also opens up a potential new source of benefit for QML: due to the potential for QML models to learn features which are classically inaccessible \cite{liu2021rigorous}, adversarial 
examples generated by attacking a classical model may struggle to fool a quantum model. This result may hold even if the features discovered by the quantum network do not provide 
any significant benefit as far as accuracy on the clean data goes; the mere fact of them being different may confer robustness against classically generated adversarial examples. Indeed, recent preliminary results have shown such characteristics for QAML networks \cite{west2022benchmarking}. Conversely, it is also an interesting open question whether quantum adversarial attacks would show the same transferability within quantum frameworks or across quantum-classical boundaries. Robustness to any adversarial examples capable of being generated by adversaries armed only with classical computers could be a powerful advantage for early adopters of QML.
\noindent
\\ \\
\noindent
\textbf{ENCODING.}
\normalsize
\noindent
Data encoding is currently one of the biggest challenges towards the practical implementation of QAML (or more generally QML) approaches \cite{larose2020robust, schuld2021supervised}, at least in the era of near-term quantum devices. The choice of data encoding has a strong influence on the classes of decision boundaries the classifier is capable of learning (and, therefore, whether it is capable of providing quantum advantage at all). Two widely adopted approaches are amplitude and phase encoding \cite{larose2020robust}, both coming with their own advantages and overheads. For example, while amplitude encoding benefits from the exponentially (in the number of qubits) large Hilbert space of the quantum computer, it also requires exponentially deep encoding quantum circuits -- a serious overhead for the implementation on near-term devices. Conversely, phase encoding, in which data is encoded into the angles of single-qubit rotations (which therefore requires the data to first have been mapped to between $0$ and $2\pi$), is more efficient with respect to circuit depth, but requires a large number of qubits, seriously limiting the ability of current quantum hardware to encode complex datasets. 
It is also possible to consider an interleaved data encoding strategy, which consists of alternating layers of data encoding and variational gates~\cite{caro2021encoding,10.21468}. Such a strategy allows for a user-controlled trade-off between the number of qubits and circuit depth, making it particularly suitable for the current NISQ era, in which hardware constraints dominate considerations of circuit architecture. Indeed, this method was employed in the recent experimental demonstration of QAML by Ren et al~\cite{ren2022experimental}.
It is also worth noting that the current QML literature is primarily based on simple datasets, with new and more efficient data encoding schemes needed for more complex datasets relevant for real-world applications. An important line of research is to either apply classical data reduction schemes such as filtering or convolutions \cite{henderson2020quanvolutional}, or using clever data compression techniques such as reported recently via a tensor network approach \cite{dilip2022data}.
\\ \\
\noindent
Resource requirement is not the only property of the classifier that is influenced by the data encoding strategy; the robustness of QVCs to various sources of noise 
have recently been explicitly linked to the chosen strategy of data encoding \cite{larose2020robust}.
While Ref. \cite{larose2020robust} considered standard noise channels, and adversarial perturbations are
a form of worst-case noise for which comparable analytical results may be less readily obtained, 
the adversarial robustness of a classifier will nonetheless
depend on the method by which the encoding is performed. Although the current strong restrictions on the number of available qubits has 
led to the use of highly efficient forms of encoding (\textit{e.g.} amplitude encoding \cite{larose2020robust}) in order to load 
complicated data into simulated quantum classifiers \cite{lu2020quantum}, as the number of 
experimentally available qubits continues to grow the choice of encoding will open up to include more alternatives and become an important 
component of any considerations of adversarial robustness. Certain forms of encoding, for example, may map any possible input into 
a well studied subset of the full Hilbert space with more desirable adversarial robustness properties \cite{liao2021robust} than 
generic amplitude encoding, which can prepare any state. 
\noindent
\\ \\
\noindent
\textbf{ARCHITECTURE.}
\normalsize
\noindent
At the heart of the standard approach to QML is an optimisable variational circuit sandwiched between a data encoding circuit and a set of measurements to determine the prediction of the classifier (see Figure~\ref{fig:1}(c)). A QVC consists of a repeated pattern of a number of parametrized single-qubit gates followed by two qubit entangling gates. The parameters of the single qubit gates are classically optimised to learn input data features, conceptually similar to the adjustment of neuron weights in classical neural networks. Despite the early successes of QVCs for simple datasets such as MNIST, there are already challenges which have started to dampen their ability to train on more complex inputs. For example, the existence of barren plateaus \cite{mcclean2018barren} in the training loss landscape poses a serious impediment to the training of large QVCs. Although it is unclear precisely what form the large-scale quantum classifiers of the future will take, it is already becoming apparent that future QML frameworks will need to employ sophisticated architectures if they are to handle more complex datasets. 
\\ \\
\noindent
Beyond QVCs, the near future may see a rise in hybrid quantum-classical models, including for example models where data is passed 
through a number of quantum filters before being passed into a standard classical neural network or convolutional neural network 
\cite{henderson2020quanvolutional}. While such models 
may have been introduced as a temporary framework within which to seek to utilise the currently limited quantum resources before the 
eventual introduction of large, fully quantum models, their long-term implications for adversarial robustness form an 
interesting line of inquiry. It may be possible, for 
example, for them to inherit robustness guarantees similar to those obtained for quantum networks~\cite{du2021quantum, weber2021optimal}, 
while also benefiting from the powerful classification capabilities of modern neural networks. Furthermore, the extent to which 
adversarial attacks directed at a classical 
neural network transfer to a quantum-classical hybrid model remains an important open question. A further interesting direction concerns the interplay of adversarial machine learning with generative adversarial networks (GANs). 
First introduced by Goodfellow \textit{et al} \cite{goodfellow2014generative}, GANs have found 
natural applications in adversarial machine learning, in both attack \cite{hu2017generating, xiao2018generating} and defence \cite{samangouei2018defense, shen2017ape}. On the offensive side, a GAN may be used to generate adversarial perturbations, and on the defensive side, given a potentially adversarial 
example, the GAN may be asked to generate an input which closely resembles it (but, hopefully, missing any adversarial 
perturbations). Quantum generalisations of GANs (qGANs) have recently been introduced and discussed in several contexts 
\cite{lloyd2018quantum, zoufal2019quantum}. 
\noindent
\\ \\
\noindent
\textbf{NOISE.}
\normalsize
\noindent
The experimental implementation of QAML faces a crucial challenge arising from the limitations of current quantum hardware, where both limited numbers of qubits and high levels of noise remain serious open problems. Although proof of concept experimental demonstrations of QML have already been reported \cite{riste2017demonstration, havlivcek2019supervised, huang2022quantum} on current quantum hardware (including a first demonstration of QAML \cite{ren2022experimental}), the sophisticated QML and QAML applications of the future will require error mitigation and correction~\cite{peters2021machine}. 
It is anticipated that for some contrived problems, quantum error mitigation may allow QAML to offer useful results in the NISQ era, however ultimately, the route to practicality of QAML approaches may go through the development of fault-tolerant quantum computers so as to accurately execute deep quantum circuits. Fault-tolerance requires that the errors induced by noise in the quantum computer be reduced beneath a certain threshold value, following which they can be systematically suppressed to arbitrarily low levels by the use of large numbers of physical qubits to encode each required logical qubit. A simplified conceptual diagram illustrating a fault-tolerant implementation of QAML over logical qubits is shown in Figure~\ref{fig:FTI} (a) which employs a surface code scheme to protect noisy qubits against errors. Surface codes are one of the most promising algorithms for the implementation of fault-tolerant quantum operations \cite{fowler2012surface}, with very recent results indicating that quantum hardware is now approaching the capability to execute small-scale implementations of surface codes \cite{acharya2022suppressing}.
\\ \\
\noindent
The estimation of exact requirements (qubit numbers, error rates, etc.) for a fault-tolerant implementation of a QAML framework is beyond the scope of this paper, however, we provide a qualitative estimate for the surface code based QAML implementation and testing which would be a first step towards practicality of QAML approaches. For this purpose, we refer to 
the IBM Quantum hardware road-map \cite{ibm_roadmap}, which indicates that in 2025 quantum devices with $\geq$4000 qubits are anticipated, with qubit numbers increasing above 10$^4$ beyond 2026. Similar road-maps have been announced by other major quantum hardware developers \cite{google_roadmap, ionq_roadmap}. 
To estimate QAML resources, we implemented (within a noiseless simulation environment) optimised 10-qubit QVCs with varying circuit depths (QVC50, QVC100, and QVC200 with 50, 100 and 200 variational layers respectively) trained on the MNIST and FMNIST datasets, with the achieved test accuracies plotted in Figure~\ref{fig:FTI} (b) and (c) \cite{west2022benchmarking}. 
These 10-qubit QML networks are quite deep, with for example QVC200 utilising about 10$^4$ quantum gates in total. 
For a fault-tolerant implementation of QVC200, we would require 10 logical qubits. Figure~\ref{fig:FTI} (d) plots the number of physical qubits required per 10 logical qubits as a function of surface code distance \cite{fowler2012surface}, indicating that distance 20 and 30 surface codes might be possible to test on IBM Quantum devices in 2025 and 2026$^+$, respectively, provided that the error rates are below the surface code threshold of about 0.5\%. Next, we estimate the number of quantum gates that can be fault-tolerantly implemented as a function of the physical error rate for various surface code distances, which to a first order approximation scales as
$n_{G} \sim (p/p_{\mathrm{th}})^{-d/2}$, where $n_G$ is the number of fault tolerant gates which can be performed, $p$ the physical error rate, 
$p_{\mathrm{th}}$ the threshold and $d$ the distance \cite{fowler2012surface}. The results are plotted in Figure~\ref{fig:FTI} (e). These plots show 
that even with an overheard of $10^2\times$ -- $10^3\times$ (required for synthesis and distillation of arbitrary rotation single qubit gates and two-qubit entangling gates \cite{mooneyquantum2021, earl2016semitech}), a surface code implementation of QVC200 of distance 20 (on IBM Quantum 2025 devices) might be possible if the error rates are around 0.15\%, rising to 0.25\% and distance 30 with 10$^4$ physical qubits. It is noted that these resource requirements are only qualitative and can be further optimised by more depth efficient variational circuits or by resource optimisation of surface code based fault-tolerant quantum gate implementations \cite{earl2021arXiv}. On the other hand, we also acknowledge that the above mentioned estimates based on 10 logical qubits are for simple image datasets only (MNIST and FMNIST), whereas complex datasets might require higher number of logical qubits or deeper quantum circuits. In summary, the continued advancements in quantum hardware have brought the implementation of QML on standard benchmark datasets to within the near-term grasp of physical devices \cite{ren2022experimental}, where their adversarial vulnerabilities will be able to be tested beyond a simulated environment. Beyond that, it can be anticipated that by the end of this decade, increasingly sophisticated experiments will be performed with advanced error correction and mitigation capabilities, expanding the range of QML and QAML to applications of practical interest.

\noindent
\\ \\
\Large{\textbf{Conclusions}}
\normalsize
\\ \\
\noindent
Quantum machine learning is generally anticipated to be one of the more widely employed use cases for quantum computers in the current NISQ period. As indicated by recently announced various quantum hardware roadmaps, increasingly capable quantum devices (larger qubit numbers and lower error rates) are expected to be deployed in the next few years, which should allow rigorous benchmarking of quantum machine learning models, likely leading to their applications for a variety of classification and feature detection problems of practical interest. Beyond that, the successful development of large-scale fault tolerant quantum 
computers may one day lead to quantum machine learning systems being entrusted with sensitive tasks which must 
be carried out with extremely high reliability. As such, and given both the existence of adversarial examples which 
can fool highly sophisticated classical neural networks and the early evidence of similar examples in the quantum setting, understanding the potential vulnerabilities of quantum classifiers, and how they compare to those of their classical counterparts, is a pressing matter which deserves further research. In this survey paper we have described the new and rapidly growing field of quantum adversarial machine learning, summarising the key achievements to date and outlining some of the main challenges that yet remain. The importance of this area is clear; if quantum machine learning is to become the revolutionary and widely applied technology it is hoped it will, then QAML must inevitably be fully understood. We anticipate that the detailed review of the field of QAML presented in our work will serve as a timely guidance for future research, and help to design robust ML tools by leveraging the unique properties of quantum computing.\\

\noindent
\textbf{Acknowledgements:} We acknowledge useful discussions with Spiro Gicev and Gary Mooney. MTW acknowledges the support of the Australian Government Research Training Program Scholarship. SME is in part supported by Australian Research Council (ARC) Discovery Early Career Researcher Award (DECRA) DE220100680. The authors acknowledge the support from Australian Army Research through Quantum Technology Challenge (QTC22). The computational resources were provided by the National Computing Infrastructure (NCI) and Pawsey Supercomputing Center through National Computational Merit Allocation Scheme (NCMAS).
\\ \\
\noindent
\textbf{Author contributions:} M.U. conceived and supervised the project. MTW and MU wrote the manuscript, with input from all authors.
\\ \\
\noindent
\textbf{Competing financial interests:} The authors declare no competing financial or non-financial interests.
 
\clearpage

\def\bibsection{\subsection*{\refname}}

\bibliographystyle{naturemag}

\end{document}